\documentclass[aps,prx,superscriptaddress,reprint,showpacs,amssymb,amsmath,floatfix,citeautoscript]{revtex4-2}

\usepackage{graphicx}
\usepackage{dcolumn}
\usepackage{bm}
\usepackage{xcolor}
\usepackage[export]{adjustbox}
\usepackage{float}
\usepackage{tabularx}
\usepackage{multirow}
\usepackage{amsmath}
\usepackage{braket}
\usepackage{array}
\usepackage{ulem}

\newcolumntype{C}[1]{>{\centering\let\newline\\\arraybackslash\hspace{0pt}}m{#1}}

\begin{document}

\preprint{APS/123-QED}

\title{Fe substitution in URu$_2$Si$_2$: Singlet magnetism in an extended Doniach phase diagram}


\author{Andrea~Marino}
\affiliation{Max Planck Institute for Chemical Physics of Solids, N{\"o}thnitzer Stra{\ss}e 40, 01187 Dresden, Germany}
\author{Denise~S.~Christovam}
\affiliation{Max Planck Institute for Chemical Physics of Solids, N{\"o}thnitzer Stra{\ss}e 40, 01187 Dresden, Germany}
\author{Chun-Fu~Chang}
\affiliation{Max Planck Institute for Chemical Physics of Solids, N{\"o}thnitzer Stra{\ss}e 40, 01187 Dresden, Germany}
\author{Johannes~Falke}
\affiliation{Max Planck Institute for Chemical Physics of Solids, N{\"o}thnitzer Stra{\ss}e 40, 01187 Dresden, Germany}
\author{Chang-Yang~Kuo}
\affiliation{Department of Electrophysics, National Yang Ming Chiao Tung University, Hsinchu 30010, Taiwan}
\affiliation{National Synchrotron Radiation Research Center, 101 Hsin-Ann Road, 30076 Hsinchu, Taiwan}
\author{Chi-Nan~Wu}
\affiliation{Max Planck Institute for Chemical Physics of Solids, N{\"o}thnitzer Stra{\ss}e 40, 01187 Dresden, Germany}
\affiliation{National Synchrotron Radiation Research Center, 101 Hsin-Ann Road, 30076 Hsinchu, Taiwan}
\author{Martin~Sundermann}
\affiliation{Max Planck Institute for Chemical Physics of Solids, N{\"o}thnitzer Stra{\ss}e 40, 01187 Dresden, Germany}
\affiliation{PETRA III, DESY, Notkestra{\ss}e 85, 22607 Hamburg, Germany}
\author{Andrea Amorese}
\altaffiliation{ASML Netherlands B.V., De Run 6501, 5504 DR, Veldhoven, The Netherlands}
\affiliation{Max Planck Institute for Chemical Physics of Solids, N{\"o}thnitzer Stra{\ss}e 40, 01187 Dresden, Germany}
\affiliation{Institute of Physics II, University of Cologne, Z\"{u}lpicher Str. 77, 50937 Cologne, Germany}
\author{Hlynur~Gretarsson}
\affiliation{PETRA III, DESY, Notkestra{\ss}e 85, 22607 Hamburg, Germany}
\author{Eric~Lee-Wong}
\affiliation{Department of NanoEngineering, University of California, San Diego, California 92093, USA}
\author{Camilla~M. Moir}
\affiliation{Department of Physics, University of California, San Diego, La Jolla, California 92093, USA}
\author{Yuhang~Deng}
\affiliation{Department of Physics, University of California, San Diego, La Jolla, California 92093, USA}
\author{M.~Brian~Maple}
\affiliation{Department of Physics, University of California, San Diego, La Jolla, California 92093, USA}
\author{Peter~Thalmeier}
\affiliation{Max Planck Institute for Chemical Physics of Solids, N{\"o}thnitzer Stra{\ss}e 40, 01187 Dresden, Germany}
\author{Liu~Hao~Tjeng}
\affiliation{Max Planck Institute for Chemical Physics of Solids, N{\"o}thnitzer Stra{\ss}e 40, 01187 Dresden, Germany}
\author{Andrea~Severing}
\affiliation{Max Planck Institute for Chemical Physics of Solids, N{\"o}thnitzer Stra{\ss}e 40, 01187 Dresden, Germany}
\affiliation{Institute of Physics II, University of Cologne, Z\"{u}lpicher Str. 77, 50937 Cologne, Germany}
\date{\today}

\begin{abstract}
    The application of pressure as well as the successive substitution of Ru with Fe in the hidden order (HO) compound URu$_2$Si$_2$ leads to the formation of the large moment antiferromagnetic phase. Here we investigate the substitution series URu$_{2-x}$Fe$_x$Si$_2$ from $x$\,=\,0.0 to 2.0 by U\,4$f$ core-level photoelectron spectroscopy and observe non-monotonic changes in the spectra. The initial increase and subsequent decrease in the spectral weight of the 4$f$ core level satellite with increasing $x$ stands for a non-monotonic 5$f$ filling across the substitution series. The competition of chemical pressure and increase of the density of states at the Fermi energy, both due to substitution of Ru with Fe, can explain such behavior. An extended Doniach phase diagram including the $x$ dependence of the density of states is proposed. Also in URu$_{2-x}$Fe$_x$Si$_2$ the ground state is a singlet or quasi-doublet state consisting of two singlets. Hence, the formation of magnetic order in the URu$_{2-x}$Fe$_x$Si$_2$ substitution series must be explained within a singlet magnetism model.
\end{abstract}

\maketitle

\section{Introduction}
The transition into an electronically ordered state at 17.5\,K in the heavy fermion compound URu$_2$Si$_2$ has attracted an enormous amount of interest since its discovery about 35 years ago\,\cite{Palstra1985,Schlabitz1986,Maple1986,Oppeneer2010,Mydosh2011,Mydosh2014,Mydosh2020}, yet the order parameter of this phase is still a matter of debate. The small antiferromagnetic ordered moment of 0.03\,$\mu_B$/U along the tetragonal $c$ axis\,\cite{Broholm1987,Niklowitz2010} is too small to account for the loss of  entropy of about 0.2\,$R$\,ln2 and changes in transport properties, so the presence of long-range magnetic order, charge density or spin density wave order has to be excluded and the name \textit{hidden order} (HO) phase was born. The possible HO multipoles of 5$f$ electrons may be classified according to the representations of U-site symmetry D$_{4h}$.  In the quasi-localized 5$f$ picture the rank-4 hexadecapole (A$_2^+$) order is a natural candidate\,\cite{Haule2009, Haule2010,Kusunose2011}; it preserves time reversal and fourfold rotational symmetry.  A superposition of a primary A$_2^+$ hexadecapole and subdominant B$_2^+$ quadrupole was proposed in\,\cite{Kung2015} with a broken fourfold symmetry. In the itinerant 5$f$ models the rank-5 dotriacontapole which breaks time reversal is preferred\,\cite{Ikeda2012,Thalmeier2014}.  Two candidates were considered: The 2-fold degenerate E$^-$ HO which breaks fourfold rotational symmetry\,\cite{Okazaki2011} and the non-degenerate A$_1^-$ which preserves this symmetry\,\cite{Kambe2018}.  All HO parameters are of the antiferromagnetic type with wave vector {\bf Q} =(0,0,1) as in the large moment antiferromagnetic phase (LMAF) that appears under pressure. Furthermore, at about 1.5\,K, URu$_2$Si$_2$ undergoes a second transition into an unconventional superconducting state. 

In heavy fermion compounds the hybridization of $f$ and conduction electrons ($f$-$d$-hybridization) plays a crucial role in the formation of the ground-state\,\cite{Floquet2005,Thalmeier2005,Coleman2007,Hilbert2007,Khomskii2010,Stockert2012,White2015,Coleman2015}. How to formulate this process in uranium compounds is, however, a subject of intense discussions. Band effects are clearly important so band structure approaches\,\cite{Oppeneer2010,Mydosh2011,Mydosh2014} have merit. Yet a localized 5$f$ electron picture may also have value\,\cite{Santini1994,Santini1998}, and recently, the existence of local atomic multiplet states was observed in U$M_2$Si$_2$ ($M$\,=\,Fe, Ni, Ru, and Pd)\,\cite{Sundermann2016, amorese2020}. The interplay with the bands is then represented by the non-integer filling of the 5$f$ shell, i.e., more than one configuration contributes to the ground state.

Motivated by the elusiveness of the HO phase, a plethora of studies explored the phase diagram of URu$_2$Si$_2$, in particular in the vicinity of the HO.  The application of pressure drives the system into the LMAFM phase. At the critical pressure of $\approx$\,5\,kbar the ordered magnetic moment rises discontinuously from 0.03 to about 0.4\,$\mu_B$\,\cite{Amitsuka1999,Bourdarot2005,Amitsuka2007,Niklowitz2010,Butch2010} and aligns along the tetragonal $c$ axis. It should be noted that a small orthorhombic distortion takes place for $p\gtrapprox$\,3\,kbar\,\cite{choi2018}. The HO phase in pristine URu$_2$Si$_2$ can also be destroyed with chemical substitution on the Ru site. Re substitution leads to ferromagnetic ordering\,\cite{dalichaouch1989}, while Rh substitution favors the LMAFM phase\,\cite{amitsuka1988, yokoyama2004, prokes2017}. Of particular interest is the isoelectronic substitution of Ru with either Fe\,\cite{das2015,williams2017, kanchanavatee2011, wilson2016, butch2016, ran2016, wolowiec2016, kung2016, ran2017,kissin2019} or Os\,\cite{wilson2016, dalichaouch1990, kanchanavatee2014, wolowiec2021}, both of which drive the system from HO into the LMAFM phase although the end members UFe$_2$Si$_2$ and UOs$_2$Si$_2$ are Pauli paramagnetic (PP) down to the lowest temperature\,\cite{Buschow1986, Palstra1986, Szytuka1988, Endstra1993, Endstra1993b}.

\begin{figure}[t!]
    \includegraphics[width=1\columnwidth]{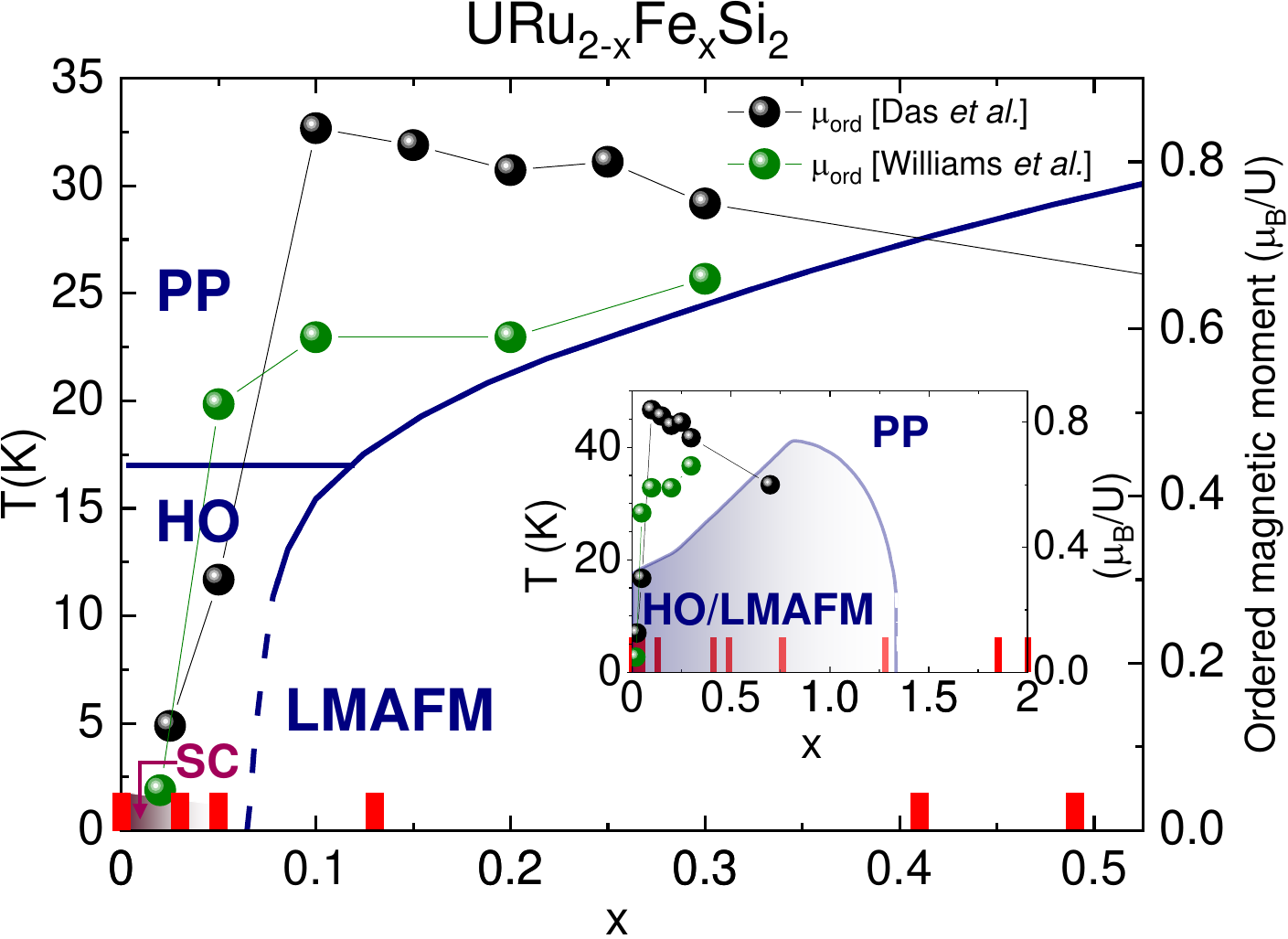}
    \caption{(left scale)\,Temperature versus $x$, where $x$ is the Fe concentration; the phase diagram is adapted from Ref.\,\cite{ran2016} and  \cite{kanchanavatee2011} showing the phase boundaries from the Pauli paramagnetic phase (PP) to the large moment antiferromagnetic (LMAFM) and superconducting (SC) phase. (Right scale) Ordered magnetic moment as measured by neutron diffraction in Ref.\,\cite{das2015, williams2017}. The red ticks mark the actual concentrations used in PES experiments. The inset shows the phase diagram and ordered moments over the full $x$ range.}
    \label{phase_diagram}
\end{figure}

Figure\,\ref{phase_diagram} shows the temperature vs Fe content phase diagram of the substitution series URu$_{2-x}$Fe$_x$Si$_2$ (lines)\,\cite{ran2016,kanchanavatee2011} along with the ordered magnetic moment (green and black dots) as a function of $x$. At first, while the moment is still increasing, the HO and LMAFM phase coexist and the magnetic volume fraction is less than 1\,\cite{wilson2016}.  The moment reaches its maximum value of about 0.8\,$\mu_B$ according to\,\cite{das2015} or  0.6\,$\mu_B$ according to\,\cite{williams2017} for $x$\,=\,0.1 with a magnetic volume of 100\%\,\cite{wilson2016}. As $x$ increases further the magnetic moment decreases whereas the N\'{e}el temperature continues to increase up to $x$\,=\,0.8 (see inset), and then quickly drops reaching zero at about  $x$\,=\,1.2\,\cite{kanchanavatee2011}. The system, for even larger $x$, enters the PP ground state of UFe$_2$Si$_2$\,\cite{Buschow1986, Palstra1986, Szytuka1988, Endstra1993, Endstra1993b}. The short tetragonal $a$ axis decreases linearly from $x$\,=\,0 to $x$\,=\,2, while the long $c$ axis remains rather constant\,\cite{kanchanavatee2011}.

The temperature versus Fe content phase diagram of URu$_{2-x}$Fe$_x$Si$_2$ bears undeniable similarities to the temperature versus pressure phase diagram of URu$_2$Si$_2$, and indeed the resemblance has been discussed for the range $x$\,=\,0\,-\,0.3 corresponding to pressures $p$\,=\,0\,-\,15\,kbar\,\cite{das2015, kanchanavatee2011}.  Furthermore, neutron scattering experiments show the same excitations both in the pressure and in the Fe substitution induced LMAFM phase\,\cite{das2015, butch2016, williams2017}, testifying to the similarity of the respective antiferromagnetic phases. In either case, the question remains open as to what causes the suppression of HO and the emergence of LMAFM. Recently, Wolowiec \textit{et al.}\,\cite{wolowiec2021} and Frantzeskakis \textit{et al.}\,\cite{Frantzeskakis2021} have pointed out the importance of changes in the 5$f$-$d$-electron hybridization for stabilizing the LMAFM phase upon the application of pressure or substitution of Fe or Os.

Photoelectron spectroscopy (PES) is the ideal tool to study hybridization effects\,\cite{Gunnarson1983, degroot1994} so that changes in 5$f$-$d$ hybridization should be observable with PES. And indeed, Fujimori \textit{et al}. have shown that the more localized compounds exhibit a stronger satellite structure in the U\,4$f$ core-level spectra\,,\cite{Fujimori1998,Fujimori2016,Fujimori2016}. This is supported by a recent core level spectroscopy study of the U$M_2$Si$_2$ family with $M$\,=\,Fe, Ru, Pd, and Ni by some of the present authors\,\cite{amorese2020} that showed that the two antiferromagnetic compounds UPd$_2$Si$_2$\,\cite{Ptasiewicz1981,Shemirani1993} and UNi$_2$Si$_2$\,\cite{Lin1991} have a much more pronounced satellite than URu$_2$Si$_2$ (HO), which in turn has a stronger satellite than the Pauli paramagnetic compound UFe$_2$Si$_2$. With a directional dependent non-resonant inelastic x-ray scattering (NIXS) experiment, analogous to linear polarization dependent x-ray absorption (XAS) in its ability to determine the ground state symmetry, it was also shown that these  members share the same ground state symmetry\,\cite{amorese2020}, namely a singlet or quasi-doublet of the U 5$f^2$ configuration so that the satellite strength in the U\,4$f$ core-level data are directly comparable and can be used to sort these members of the U$M_2$Si$_2$ ($M$\,=\,Fe, Ni, Ru, and Pd) family into a Doniach phase diagram, temperature $T$ versus exchange interaction $\cal{J}$, that describes the competition between magnetic order and a non-magnetic delocalized state of the $f$ electrons. For small $\cal{J}$, magnetic order prevails, while for large values of $\cal{J}$, the itinerant character of the ground state dominated. In-between these two regimes, a quantum critical point occurs and, in its vicinity, superconductivity often occurs.  URu$_2$Si$_2$, which is also superconducting at low $T$, is placed in the vicinity of the quantum critical point, the Fe compound on the side of larger $\cal{J}$ in the Pauli paramagnetic regime, and the Ni and Pd compounds at smaller values of $\cal{J}$ with respect to URu$_2$Si$_2$, well in the magnetic ordering part of the Doniach phase diagram. 

The present study focuses on the vicinity of the HO phase in URu$_2$Si$_2$. The substitution series URu$_{2-x}$Fe$_x$Si$_2$ is investigated with core-level PES from the HO, to the LMAFM and PP phase. URu$_{2-x}$Fe$_x$Si$_2$ samples with nominal composition $x$ that cover the entire phase diagram were measured. In Figure\,\ref{phase_diagram} the \textit{actual} compositions (see below) of the measured samples are marked by red ticks.  We also present directional dependent NIXS data for single crystalline URu$_{1.7}$Fe$_{0.3}$Si$_2$,, complementing the NIXS spectra of URu$_2$Si$_2$ and UFe$_2$Si$_2$ of Refs.\,\cite{Sundermann2016, amorese2020}.

\section{Experiment}
Polycrystalline samples of Fe-substituted URu$_2$Si$_2$ used in the PES experiment were prepared by arc-melting in an argon atmosphere and then analyzed by x-ray powder diffraction to determine quality and stoichiometry.  The actual Fe concentration, as examined by elemental analysis using energy dispersive x-ray spectroscopy, is uniform throughout the sample and in a good agreement with the nominal concentration.  
The actual relative composition of the spots measured with PES was also checked by comparing the intensities of the U\,4$d$ and Fe\,2$p$ core level emission lines and was very much in agreement with the nominal composition (see Tab.\,\ref{tab:compostion}).

Single crystalline URu$_{1.7}$Fe$_{0.3}$Si$_2$ used in the NIXS experiment was grown by the Czochralski method in a tetra-arc furnace from high purity starting elements. Pieces are taken from the single crystalline parts of the boule and a small piece of the boule is then analyzed \cite{butch2016} by powder diffraction to determine structure and quality. The real Fe concentration was examined by elemental analysis using energy dispersive x-ray spectroscopy, which is uniform throughout the sample and in a good agreement with the nominal concentration.

Soft x-ray PES experiments were performed at the NSRRC-MPI TPS 45A Submicron Soft X-ray Spectroscopy beamline\,\cite{huangming2019} at the Taiwan Photon Source in Taiwan. The incident photon energy was set to 1200\,eV. The valence band spectrum of a platinum sample was measured in order to determine the chemical potential and the overall instrumental resolution of about 200\,meV. The excited photoelectrons were collected using a MB Scientific A-1 photoelectron analyzer in the horizontal plane at 60$^{\circ}$. The sample was measured at normal emission. Clean sample surfaces were obtained by cleaving the samples \textsl{in situ} in the cleaving chamber prior to inserting them into the main chamber. In both chambers the pressure was in the low 10$^{-10}$\,mbar regime. The measurements were performed at a temperature of 40\,K. 

The NIXS experiment was performed at the High-Resolution Dynamics Beamline P01 of the PETRA-III synchrotron in Hamburg, Germany. The sample temperature was kept at 15\,K in a vacuum cryostat. The incident energy was selected with a Si(311) double monochromator. The P01 NIXS end station has a vertical geometry with twelve Si(660) 1 m radius spherically bent crystal analyzers, fixing the final energy to 9690 eV. The analyzers are position at $2\theta\approx 155^o$, giving a momentum transfer of $|\mathbf{q}|\approx9.6$\AA$^{-1}$. The experimental energy resolution, determined by regularly measuring the elastic line, amounts to $\approx0.7$\,eV. More details on the setup can be found in Ref.\,\cite{Sundermann2017}. 
The multipole selection rules in a NIXS experiment at high momentum transfers $q$ give access to the ground state symmetry in a similar fashion as the dipole selection rules in x-ray absorption spectroscopy  XAS\,\cite{Schuelke2007,Haverkort2007}. In NIXS, the direction of the momentum transfer, i.e., $\vec{q}$, takes over the role of the electric field vector $\vec{E}$ in XAS. For uranium compounds, high $q$ NIXS has the advantage of more excitonic excitations and the non-resonant process facilitates the quantitative modeling\,\cite{Caciuffo2010,Sundermann2016,Sundermann2018a,amorese2020}.

\section{Results}

\begin{figure}[t!]
    \includegraphics[width=0.9\columnwidth]{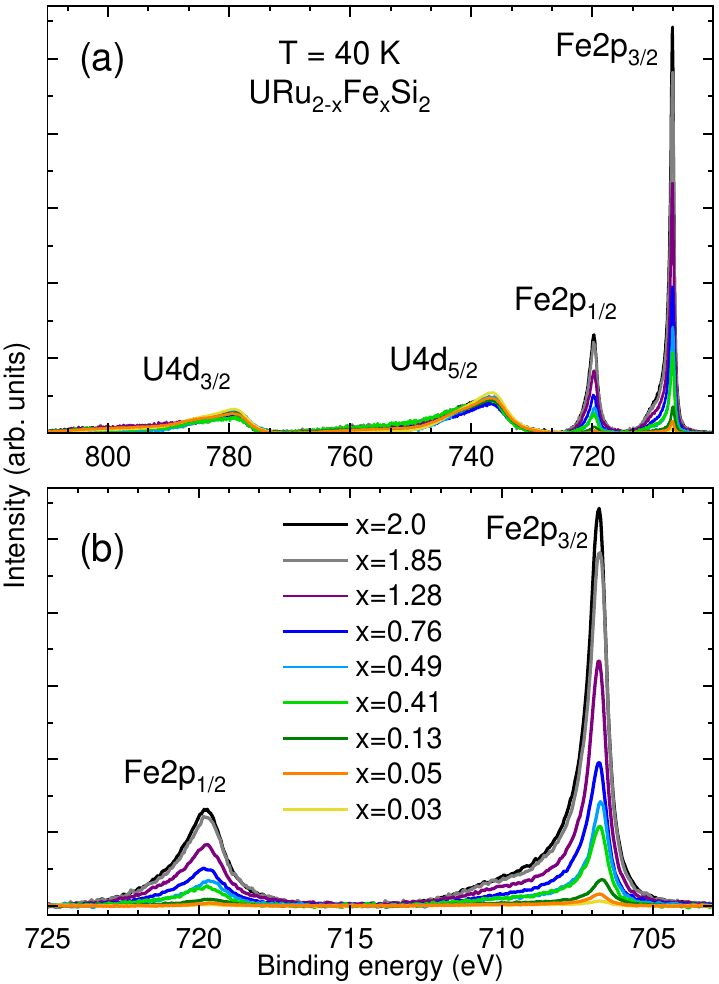}
    \caption{(a) U\,$4d$ and Fe\,$2p$ and (b) Fe\,$2p$ core level spectra of URu$_{2-x}$Fe$_x$Si$_2$. The spectra are integrated background (Shirley) corrected and normalized to the area of the U\,$4d$ peaks.}
    \label{U4dFe2p}
\end{figure}

\begin{figure}[t]
    \includegraphics[width=1\columnwidth]{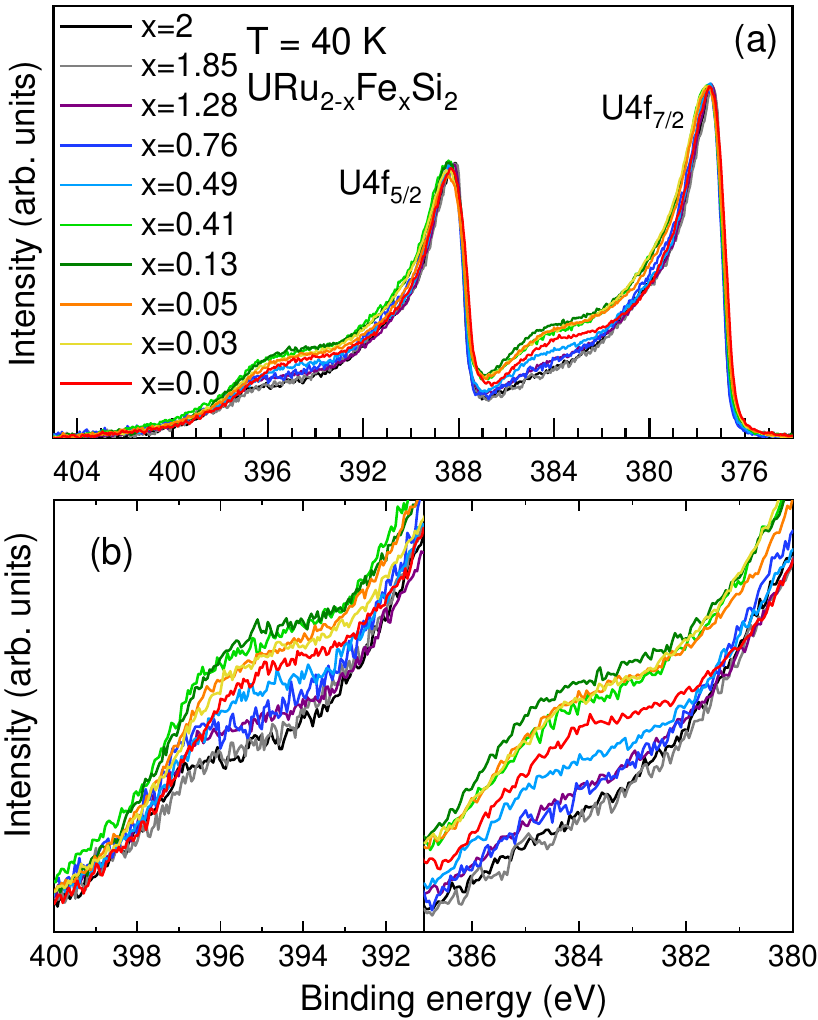}
    \caption{(a) U\,4$f$ core level data of URu$_{2-x}$Fe$_x$Si$_2$. The spectra are normalized to the main peak of the spectrum and an integrated background (Shirley) is subtracted. (b) Blow-up on the satellite regions of the U\,4$f_{5/2}$ (left) and U\,4$f_{7/2}$ (right) lines. }
    \label{4f}
\end{figure}

Figure\,\ref{U4dFe2p}\,(a) shows the U\,$4d$ and Fe\,$2p$ core level spectra of URu$_{2-x}$Fe$_x$Si$_2$ after the subtraction of an integrated (Shirley) background and normalization to the integrated intensity of the U\,$4d$ lines. The U\,$4d$ can be used for normalization because it is a full shell so that the PES intensities should not vary across the URu$_{2-x}$Fe$_x$Si$_2$ series. The Fe\,$2p$ is also a full shell, so that differences in its intensity must be due to the different Fe content in the samples, making it suitable for determining the actual compositions. Figure\,\ref{U4dFe2p}\,(b) shows a blow-up of the Fe\,$2p$ lines. We consider the ratios of the integrated Fe\,$2p$ area of a sample with Fe content $x$ to the integrated Fe\,$2p$ area of the UFe$_2$Si$_2$ sample. We take the latter as reference. The $x$\,=\,0 sample does not show any Fe line. We then compare the ratios to their nominal values and make an estimate of the composition. We find that the composition of the spots measured with PES agree well with the nominal compositions, as shown in Table\,\ref{tab:compostion}. Only the nominal $x$\,=\,0.08 sample seems to have a slightly lower actual composition so that also this sample is still in the HO phase. In the following we keep referring to the actual composition. 
\newcolumntype{Y}{>{\centering\arraybackslash}X}
\begin{table}[]
\begin{minipage}[t]{0.49\columnwidth}
    \noindent
    \flushleft
    \begin{tabularx}{1\columnwidth}{| Y | Y |}
        \hline
        \small{Nominal composition} &  \small{Actual composition} \\ \hline \hline
        \small{0.00} & \small{0.00} \tiny{HO} \\      
        \small{0.05} & \small{0.05} \tiny{HO/AFM}\\
        \small{0.08} & \small{0.03} \tiny{HO/AFM}\\ 
        \small{0.1} & \small{0.13} \tiny{AFM}\\
        \small{0.4} & \small{0.41} \tiny{AFM}\\
        \hline
    \end{tabularx}
\end{minipage}
\hfill
\begin{minipage}[t]{0.49\columnwidth}
    \noindent
    \flushright
    \begin{tabularx}{1\columnwidth}{| Y | Y |}
        \hline
        \small{Nominal composition} &  \small{Actual composition} \\ \hline \hline
        \small{0.5} & \small{0.49} \tiny{AFM}\\
        \small{0.8} & \small{0.76} \tiny{AFM}\\ 
        \small{1.3} & \small{1.28} \tiny{AFM/PP}\\
        \small{1.8} & \small{1.85} \tiny{PP}\\
        \small{2.0} & \small{2.00} \tiny{PP}\\
        \hline
    \end{tabularx}
\end{minipage}
\caption{Nominal composition compared to the composition estimated from the Fe\,$2p$ intensity, as normalized to the U\,$4d$ integrated intensity. The Fe\,2$p$ intensity for $x$\,=\,2 serves as the reference composition. The labels indicate the phase: HO is hidden order, AFM is antiferromagnetic order and PP is Pauli paramagnetic.}
\label{tab:compostion}
\end{table}
Figure\,\ref{4f}\,(a) shows the U\,4$f$ core-level spectra of the URu$_{2-x}$Fe$_x$Si$_2$ samples after the subtraction of an integrated (Shirley) background and normalization to the U\,4$f_{7/2}$ peak at $\approx$\,377.4\,eV binding energy. The U\,4$f_{5/2}$ and U\,4$f_{7/2}$ multiplet structures are split by about 11\,eV by the U\,4$f$ spin orbit interaction. Both the U\,4$f_{5/2}$ and U\,4$f_{7/2}$ emission lines originate from the same mixed valent initial state U configuration and they both show satellite structures. The ratio of the integrated intensities of the U\,4$f_{5/2}$ and U\,4$f_{7/2}$ emission lines (main and satellite) is about 0.8 which is close to the expectation value 6/8\,=\,0.75.

We are searching for changes of spectral weights as a function of substitution rather than attempting to determine the absolute occupation of the 5$f$ shell. We therefore focus on the satellite structure, as shown in Figure\,\ref{4f}\,(b) for the U\,4$f_{5/2}$ (left) and U\,4$f_{7/2}$ (right) emission lines. The $x$\,=\,0 (URu$_2$Si$_2$) satellite structure in the HO phase is much more pronounced than the one of the $x$\,=\,2 sample (UFe$_2$Si$_2$) in the PP phase, in agreement with previous findings\,\cite{amorese2020}. Looking at the different compositions in detail we find that starting from $x$\,=\,0, the satellite at first increases, peaks for $x$\,=\,0.13, then decreases, and becomes smaller than for URu$_2$Si$_2$ for $x$\,$\geq$\,0.49 until it smoothly overlaps with the $x$\,=\,2.0 spectral shape. The satellite structures of both emission lines consistently show the same trend.

\begin{figure}[t!]
    \includegraphics[width=0.9\columnwidth]{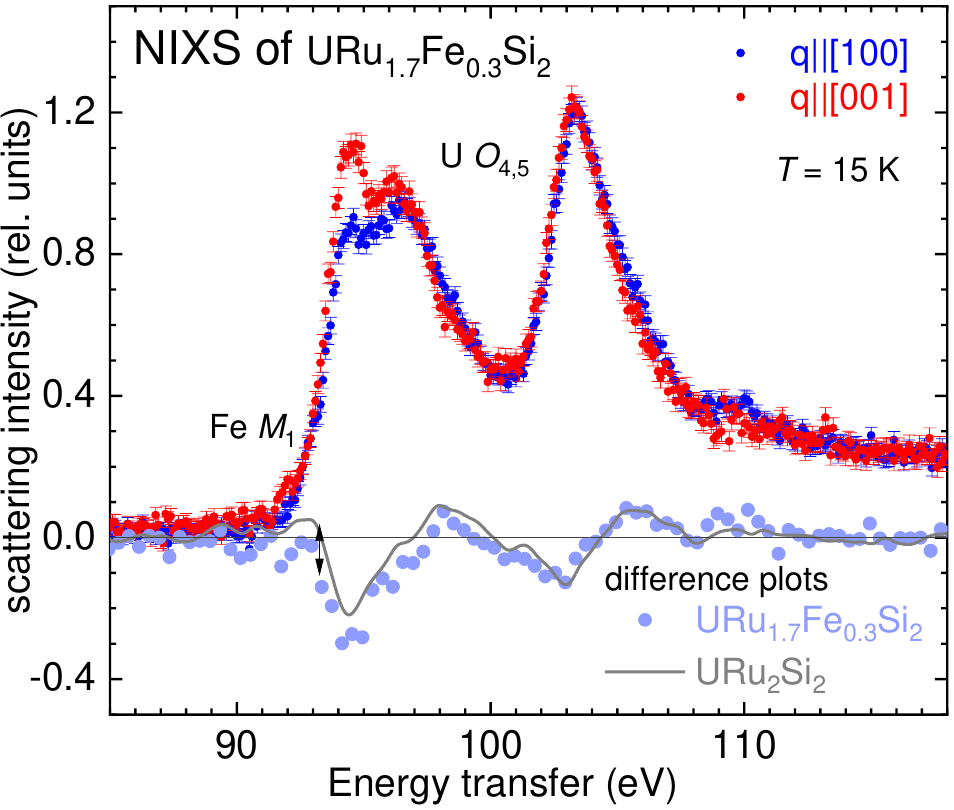}
    \caption{Normalized and background corrected experimental NIXS data of URu$_{1.7}$Fe$_{0.3}$Si$_2$ at the U $O_{4,5}$ edges (5$d$\,$\rightarrow$\,5$f$) at $T$\,=\,15\,K for $\vec{q}$$\|$[100] (blue dots) and  $\vec{q}$$\|$[001] (red dots), plus the difference plots I$_{\vec{q}\|[100]}$-I$_{\vec{q}\|[001]}$ (violet dots). The arrow marks the region where Fe $M_1$ scattering appears in the spectra. For comparison the difference data of URu$_2$Si$_2$ are shown (gray line), adapted from Ref.\,\cite{Sundermann2016}. When no error bars are given the size of the data points represent the statistical error.}
    \label{fig_NIXS}
\end{figure}

Before discussing the above observation further, it should be emphasized that the comparison of PES spectra in terms of hybridization is only sensible if other changes in, e.g., ground state symmetry and/or degeneracy can be excluded. Indeed, we already know from Ref.\,\cite{Sundermann2016,amorese2020} that the end members of the substitution series URu$_2$Si$_2$ and UFe$_2$Si$_2$ share the same symmetry, so it is reasonable to assume this is also the case for the in-between compositions. For confirmation, Figure\,\ref{fig_NIXS} shows the U $O_{4,5}$ edge (5$d$\,$\rightarrow$\,5$f$) NIXS data of URu$_{1.7}$Fe$_{0.3}$Si$_2$ at 15\,K for two directions of the momentum transfer $\vec{q}$, for $\vec{q}$$\|$[100] (blue) and $\vec{q}$$\|$[001] (red).  Two aspects are striking, namely the strong directional dependence and the existence of a multiplet structure. The spectral shape and directional dependence is the same as for URu$_2$Si$_2$ and UFe$_2$Si$_2$. These data show that the ground state symmetry remains the singlet $\Gamma_1^{(1)}(\theta) =  \frac{1}{\sqrt{2}} \sin\theta ( | +4 \rangle + | -4 \rangle )+\frac{1}{\sqrt{2}} \cos\theta|0\rangle$ or the $\Gamma_2 =\frac{1}{\sqrt{2}}  ( | +4 \rangle - | -4 \rangle )$, or a quasi-doublet constructed from these two singlet states also for this intermediate Fe concentration.

\section{Discussion}
Uranium in intermetallic compounds is usually strongly intermediate valent with a valence between 4+ and 3+, i.e., with a configuration between 5$f^2$ and 5$f^3$. In previous works, it was shown that the symmetry of URu$_2$Si$_2$ is determined by the U 5$f^2$ configuration, implying this is the configuration lowest in energy and that the $f$-level filling is closer to 2 than to 3.  The degree of itinerancy can be identified as the deviation $n_f$ from integer valence ($f^2$) or, in other words, it is given by the extra participation of a third electron when following the \textit{dual-nature} concept of $f$ electrons as discussed for some U intermetallic compounds \,\cite{Thalmeier2002,zwicknagl2003a,Lee2018}. Strong satellites in the 4$f$ core level PES spectra of uranium inter-metallic compounds are interpreted as a sign of strong localization\,\cite{Fujimori1998,Fujimori2016,Fujimori2016,amorese2020}. Hence, within the U$M_2$Si$_2$ series where the satellites appear at the same binding energy, the satellite spectral weight is related to the occupation of the 5$f$ shell, whereby strong satellites stand for more $f^2$ and favor magnetism, while weaker satellites imply more itinerancy due a larger contribution of a third electron ($f^3$ configuration) and a Pauli paramagnetic ground state. 

\begin{figure}[t!]
    \includegraphics[width=1\columnwidth]{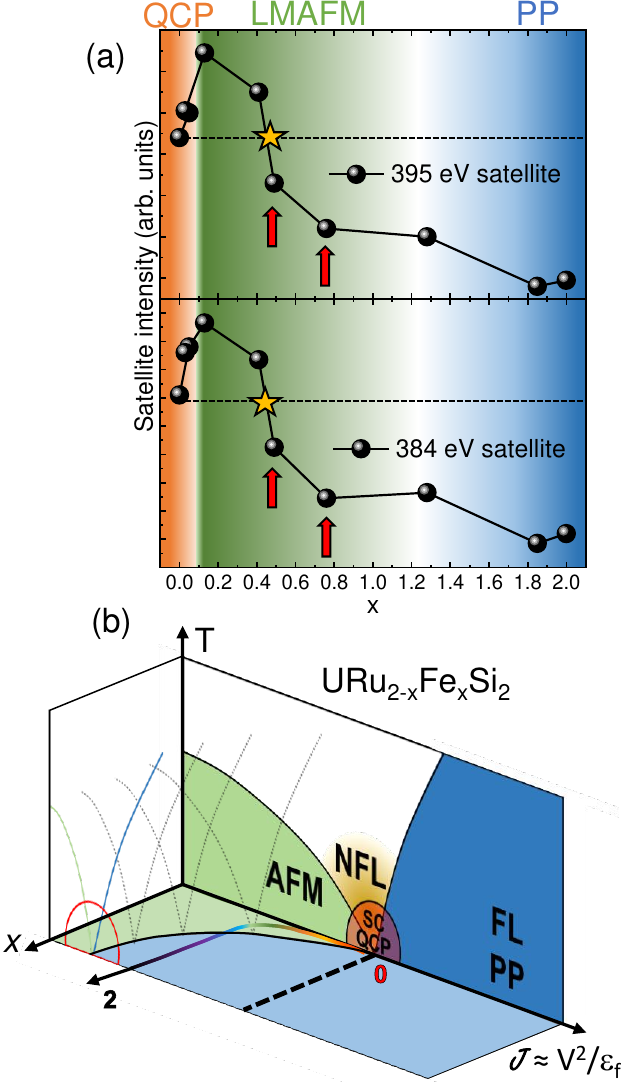}
    \caption{(a) Satellite intensities of the U\,$4f$ emission spectra at 395\,eV and 384\,eV binding energy as a function of $x$. The red arrows indicate values of $x$ showing a smaller satellite structure than $x=0$, despite being in the LMAFM phase.(b) Extended Doniach phase diagram ($T$,$\cal{J}$,$x$) of URu$_{2-x}$Fe$_x$Si$_2$. LM stands for large moment, AFM for antiferromagnetic, SC for superconducting, QCP for quantum critical point, NFL for Non-Fermi liquid and FL for Fermi liquid, and PP Pauli paramagnetic. }
    \label{Fig_Doniach}
\end{figure}

In the URu$_{2-x}$Fe$_x$Si$_2$ series, the satellite spectral weight at first increases up to $x$\,=\,0.13, then decreases again, and becomes smaller in comparison to URu$_2$Si$_2$ for $x$\,$\ge$\,0.49. This alternating trend in the core level intensities (or $f$ electron count) comes as a surprise because there are no anomalies in, e.g., the lattice constants or cell volume as function of $x$ in URu$_{2-x}$Fe$_x$Si$_2$\,\cite{kanchanavatee2011}. Moreover, the satellite intensities for 0\,$<$\,$x$\,$<$\,0.4 cannot be composed of the satellite strengths of the end members of the series. This becomes evident when looking at Figure\,\ref{Fig_Doniach}\,(a) where the intensities of the U\,4$f$ spectra at 395 and 384\,eV binding energy are plotted as a function of the Fe concentration $x$. The background colors indicate the respective phases, i.e., the hidden order phase close to a quantum critical point (QCP) and region of phase coexistence in orange, the magnetic (LMAFM) region in green, and the Pauli paramagmetic (PP) phase in blue. The horizontal dashed lines mark the satellite spectral weight of URu$_2$Si$_2$. 
\subsection{The effect of chemical pressure}
The substitution of Fe in URu$_2$Si$_2$ leads to a continuous decrease of the unit cell volume so that, like applied pressure, the configuration with the smaller ionic radius, the U\,$f^2$, becomes  the more favourable one. This suggests the satellite intensity will increase. And indeed, the satellite intensity is at first more pronounced, thus proving that the U\,$f^2$ configuration is stabilized. However, by reducing the lattice spacings due to chemical pressure, the hybridization $V$ of $f$ and conduction electron states will also increase. Hence, when keeping in mind that a Kondo state is nearby, the material can only reach the LMAFM phase when the on-site exchange interaction 
\begin{eqnarray}
\cal{J}& \propto& \frac{V^2} {|\epsilon_f|},  
\end{eqnarray}
with $\epsilon_f$ the $f$-level position relative to the Fermi energy\,\cite{Schrieffer1966,Coleman2015}, becomes smaller with chemical pressure. Here it is assumed that URu$_2$Si$_2$, being superconducting at low temperatures, is close to the QCP so that the AFM region can only be reached with pressure when $\cal{J}$ decreases. This, in turn, can only happen when the parameter $|\epsilon_f|$ increases faster with pressure than $V^2$. This situation is reminiscent of Yb heavy fermion compounds where pressure stabilizes the smaller $f^{13}$ configuration, i.e., makes Yb more magnetic at the detriment of itinerancy, although $V$ increases (see e.g.\,\cite{Okamura2019}). 

So far our findings support the similarities of the ($p$,$T$) phase diagram of URu$_2$Si$_2$ for pressures $\le$\,15\,kbar and the ($x$,$T$) phase diagram of URu$_{2-x}$Fe$_x$Si$_2$ for x\,$\le$\,0.3\,\cite{kanchanavatee2011,das2015}, and are in agreement with the additivity of chemical and external pressure as suggested in Refs.\,\cite{kanchanavatee2011, wolowiec2016}. Also, Das \textit{et al.}\,\cite{das2015} and Kanchanavatee \textit{et al.}\,\cite{kanchanavatee2011} discuss reaching the LMAFM phase in terms of increasing $V^2$, but without considering the necessity of $\epsilon_f$ in the presence of a nearby Kondo state.

\subsection{The additional effect of substitution}
The satellite spectral weight decreases with further increase of $x$ and eventually becomes smaller than in URu$_2$Si$_2$. This could suggest that now $\cal{J}$ is increasing with $x$. However, URu$_{2-x}$Fe$_x$Si$_2$ is still magnetic despite the smaller satellite between $x\approx0.4$ and $x\approx1.2$ (see red arrows in Figure\,\ref{Fig_Doniach}\,(a)). If the satellite strength, and with it the $f$ electron count, were only driven by $\cal{J}$, we would then rather expect a QCP where the satellite spectral weight of URu$_2$Si$_2$ is recovered (see yellow star in Figure\,\ref{Fig_Doniach}\,(a)). This satellite strength would denote the quantum critical value of the exchange interaction $\cal{J}$$_c$. This is however not the case: the yellow star still lies in the LMAFM regime. We would further expect a Pauli paramagnetic regime for the entire region with satellites smaller than in URu$_2$Si$_2$, i.e., also for the concentrations marked by the red arrows in Figure\,\ref{Fig_Doniach}\,(a).  A dilemma is then posed. It cannot be solved even if we were to further speculate that $\cal{J}$ itself has a non-monotonic behaviour: even if $\cal{J}$ increased, after its decrease at low $x$ due to chemical pressure, we would never expect antiferromagnetic order to occur for any Fe concentration with a smaller satellite than pure URu$_2$Si$_2$. We take this as an indication that something else, in addition to chemical pressure, must drive the ground state properties and $f$-electron count.  

\begin{figure}[t!]
    \includegraphics[width=0.90\columnwidth]{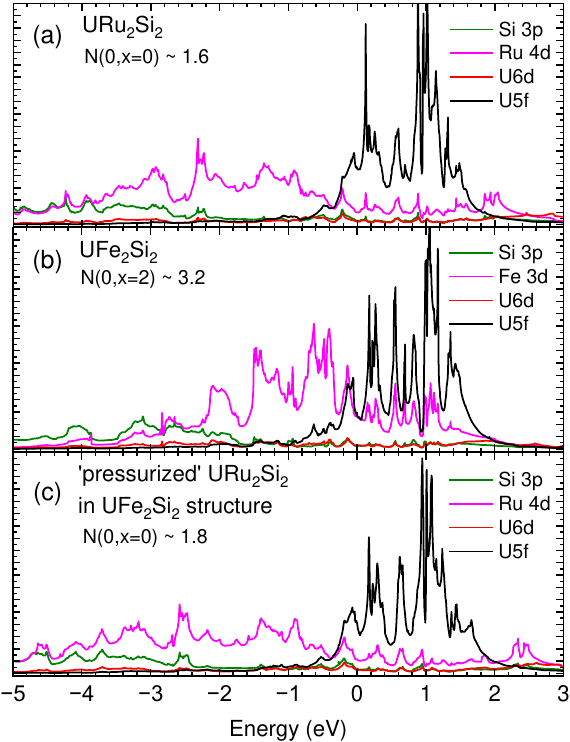}
    \caption{ Partial densities of states (p-DOS) of (a) URu$_2$Si$_2$, (b) UFe$_2$Si$_2$ and (c) URu$_2$Si$_2$ in the structure of UFe$_2$Si$_2$ to mimic the effect of pressure. $N(0,x)$, x= Fe content, is the sum of the p-DOS of the conduction electron states at the Fermi energy $E_F$, averaged on a 25\,meV interval around $E_F$.}
    \label{DOS}
\end{figure}

Figure\,\ref{DOS}\,(a) and (b) shows the partial density of states (p-DOS) of URu$_2$Si$_2$ and of UFe$_2$Si$_2$. The calculations are performed with the full-potential local-orbital (FPLO) code, employing the local density approximation (LDA) and in a full relativistic approach. A grid of 18x18x18 k-points and about one energy point every 8\,meV is used in the calculation of the p-DOS. The magenta lines represent the transition metal $d$ states. It is evident that the p-DOS has changed significantly from Ru to Fe despite the isoelectronic character of the substitution; $N(0,x)$, the sum of the p-DOS of the conduction electron states at the Fermi energy, has doubled (see Figure\,\ref{DOS}) from x\,=\,0 to x\,=\,2. In contrast, the p-DOS of URu$_2$Si$_2$ is hardly affected by pressure as the comparison with panel (c) of Figure\,\ref{DOS} confirms. It shows the same calculation as in panel (a) but with the lattice parameters of UFe$_2$Si$_2$ in order to mimic the pressure effect only. For URu$_2$Si$_2$ with pressure, $N(0)$ has hardly changed with respect to the value for ambient pressure URu$_2$Si$_2$. Hence, substitution has the additional effect of changing the density of states at the Fermi level, which we express in terms of the number $N(0,x)$.  

The density of states at the Fermi level enters the expression of the Kondo temperature 
\begin{eqnarray}
T_K & \propto & \exp{\left(-\frac{1}{N_f N(0,x)\cal{J}}\right)}.
\end{eqnarray}
Here $T_K$ is defined as in the large $N_f$ expansion\,\cite{Bickers1987,Newns1987} with $N_f$\,=\,2, the degeneracy  of the quasi-doublet ground state provided that the splitting of the latter is smaller than $T_K$.
$T_K$ decreases with decreasing $\cal{J}$ but it increases with increasing $N(0,x)$. And indeed, for large $x$ the material is Pauli paramagnetic showing that $T_K$, due to the now dominating impact of $N(0,x)$, increases so much that the magnetic order is suppressed. A more dominating Kondo regime with respect to the magnetic regime suggests that the critical exchange interaction $\mathcal{J}_c$ at the QCP has moved to smaller values. $\mathcal{J}_c$ is now also a function of the Fe concentration $x$.  Hence, the standard Doniach phase diagram is no longer sufficient. It now requires $x$ as a third dimension (see Figure\,\ref{Fig_Doniach}\,(b)).  With a simple \textit{heuristic} argument it can be shown that the QCP moves on a concave line as shown by the black line in the ($x$,$\cal{J}$)-plane in Figure\,\ref{Fig_Doniach}\,(b) (see Appendix\,\ref{app:heuristic}). The coloured line in Figure\,\ref{Fig_Doniach}\,(b) indicates a possible path for URu$_{2-x}$Fe$_x$Si$_2$ across the phase diagram, given that, as reasoned above, the system is first antiferromagnetic and eventually Pauli paramagnetic. 

What do the above considerations imply for the filling of the 5$f$ shell that we see materialized in the satellite strength? Following, e.g., Fulde\,\cite{fulde1995} $n_f$, expressed as the ratio of $T_{K}$ and $T_{RKKY}$, represents the competing effects of pressure and density of states. Here, we recall:
\begin{eqnarray}
T_{RKKY} & \propto & {\cal{J}}^2 N(0,x)  .
\end{eqnarray}
For small $x$, the $N(0,x)$ effect is not yet global, because not every U atom has an Fe atom as nearest neighbour. Here, pressure and the consequent decrease of $\cal{J}$ are the dominant effects, stabilizing the $f^2$ configuration and decreasing $n_f$ (satellite increases).  Eventually, for $x$\,$>$\,1.2, i.e., when the Pauli paramagnetic regime is reached, $n_f$ has indeed increased (satellite decreased) because now the density of states effect acts on all U atoms, thus making the Kondo effect overrule the magnetic ordering. The peculiar behaviour in the interim section with the red arrows (Figure\,\ref{Fig_Doniach}\,(a)) may show that the pressure effect is still not negligible with respect to the growing density of states effect. It could also be due to the fact that the N\'{e}el temperature is still increasing up to $x$\,$\approx$\,1.

\subsection{Induced magnetism in the presence of a singlet ground state}
\label{appendix:singlet_magnetism}
It has not yet been discussed why the N\'{e}el temperature increases up to $x$\,$\approx$\,1 despite the decrease of the magnetic moments beyond $x$\,=\,0.1\,\cite{das2015}. This brings up the question of how magnetic order can form at all in the presence of a singlet or quasi-doublet state consisting of the two singlet states $\Gamma_1^{(1)}(\theta) =  \frac{1}{\sqrt{2}} \sin\theta ( | +4 \rangle + | -4 \rangle )+\frac{1}{\sqrt{2}} \cos\theta|0\rangle$ and $\Gamma_2 =\frac{1}{\sqrt{2}}  ( | +4 \rangle - | -4 \rangle )$, neither of which carrying a local magnetic moment.  It should be mentioned that the dynamical mean field theory (DMFT) calculations by Haule and Kotliar find the same states to have the largest weight in the DMFT density matrix\,\cite{Haule2009,Haule2010}. The authors of the Ref.\,\cite{Haule2009,Haule2010} further state that these two states are compatible with antiferromagnetic order as well as hexadecapole order. Hence, the HO--LMAFM instability upon application of pressure or substitution may also be seen as the competition of these two order parameters.

\begin{figure}[t!]
    \includegraphics[width=0.9\columnwidth]{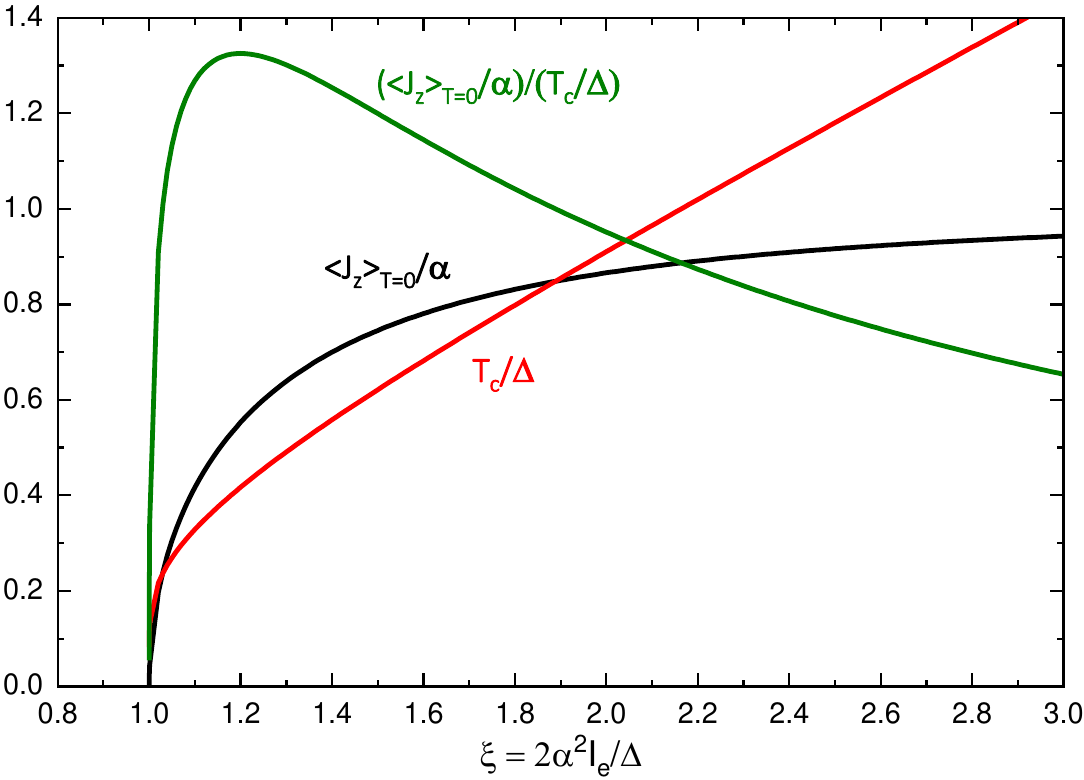}
    \caption{Normalized saturation $\langle J_z \rangle_{T=0}$ and ordering temperature $T_C$ as function of the control parameter $\xi$ for a singlet--singlet system.}
    \label{fig_induced}
\end{figure}

A moment in the presence of a quasi-doublet consisting of two singlet states can only be acquired spontaneously below the ordering or critical temperature $T_c$ by coupling these two singlet states via intersite exchange $I_e\sim T_{RKKY}$, i.e., for \textit{induced order} to occur there has to be a non-zero matrix element $\alpha=\langle gs|J_z|es\rangle$ between the singlet ground state $|gs\rangle$ and the excited state $|es\rangle$. The material orders magnetically below $T_c$ when the control parameter $\xi$ is larger than 1\,\cite{Thalmeier2002, thalmeier2021}, whereby $\xi$ depends on the ratio of the intersite exchange interaction $I_e$ and the energy difference of the two singlet states $\Delta$, times the square of the matrix element $\alpha$ (see Appendix\,\ref{app:induced}). For $\xi$\,$>$\,1, the quasi-doublet $\Gamma_1^{(1)}$-$\Gamma_2$ may carry any magnetic moment between  0 and 4\,$g_J\mu_B$ along the tetragonal $c$ axis.  Moreover, it is natural for induced moment magnetism that the ordered moment saturates while the corresponding critical temperature still increases (see Fig.\,\ref{phase_diagram}) in the limit of large control parameters where they asymptotically approach the conventional behavior of a magnet with a Kramers doublet ground state. In order to illustrate such a trend, we show in Figure\,\ref{fig_induced} the magnetic moment $\langle J_z \rangle _{T=0}/\alpha$, the critical temperature $T_c/\Delta$, and their ratio $(\langle J_z \rangle _{T=0}/\alpha)/ (T_c/\Delta)$ as a function of the control parameter $\xi$ in the singlet-singlet model. This dependence is reminiscent of the $T$-$x$ phase diagram of URu$_{2-x}$Fe$_x$Si$_2$ in the low concentration regime where $\mu_{ord}$ reaches its maximum value while $T_N$ still inscreases (see Figure\,\ref{phase_diagram}). The presence of a sufficiently strong Kondo-type interaction will destroy such induced order, i.e., the Doniach concept is also applicable for induced order. 

The induced character of the magnetic order introduces two more parameters, namely, the dipolar matrix element $\alpha$ and the energy splitting $\Delta$ between the two singlet states forming the quasi-doublet, that may change with the Fe concentration $x$. This may contribute to the formation of magnetic order despite the larger degree of itinerancy in the interim concentration regime.  

\section{Concluding remarks}

The substitution series URu$_{2-x}$Fe$_x$Si$_2$ has been investigated with 4$f$ core-level photoelectron spectroscopy (PES) and non-monotonic shifts of spectral weight have been observed. The initial increase and subsequent decrease of the satellite intensities of the U\,4$f$ core level with $x$ are interpreted in terms of two competing effects; chemical pressure stabilizing the $f^2$ configuration and hence magnetic order and the increase of the density of states at the Fermi energy enhancing the number of itinerant electrons. This has been captured in an extended Doniach phase diagram with a ($\cal{J}$,$x$)-plane in addition to the standard ($T,\cal{J}$) plane and a quantum critical point moving as function of the Fe concentration $x$. The discussion in terms of a Doniach phase diagram is possible because it was confirmed that URu$_2$Si$_2$, UFe$_2$Si$_2$ and also the substitution series share the same ground state symmetry, namely a singlet or quasi-doublet state. Finally, the formation of magnetic order has been discussed in terms of singlet magnetism.

\section{Appendix}

\subsection{Doniach phase diagram heuristics}
\label{app:heuristic}
The relevant scales within the Doniach phase diagram approach are the Kondo and RKKY interactions, whose characteristic energy scales are respectively given by:

\begin{equation}
    T_K (x) = \frac{1}{\rho(x)}\text{exp}\left(-\frac{1}{N_f \rho (x) \cal{J}\text{($x$)}}\right )
\end{equation}
\begin{equation}
    T_{RKKY} (x) = \cal{J}\text{($x$)} ^{\text{2}} \rho\text{($x$)}
\end{equation}
where $x$ is the Fe concentration, $N_f=2$ is the degeneracy of the quasi-doublet ground state, $\rho(x)=N(0,x)$ is the density of conduction states at the Fermi level and  $\cal{J}\text{($x$)}$ is the on-site exchange coupling. Let us define:

\begin{equation}
    \tilde{g}(x) = \cal{J}\text{($x$)} \rho\text{($x$)}
\end{equation}
At the quantum critical point the condition $T_{RKKY}=T_K$ holds, which means in terms of the quantum critical $\tilde{g}_c(x)$: 

\begin{equation}
    \tilde{g}_c(x)^2=exp \left( -\frac{1}{2\tilde{g}_c(x)}\right)
\end{equation}
This equation has the same solution $\tilde{g}(x)= \text{const} (=0.12 \ll 1)$ for all $x$. In other words:

\begin{eqnarray}
    \tilde{g}_c(x)=\text{const} = \tilde{g}_c(0) \\
    \cal{J}\text{$_c(x$)} \rho\text{$_c(x$)} = \cal{J}\text{(0)} \rho\text{(0)} \label{line}
\end{eqnarray}
From Eq.\,\ref{line} we can see that $\mathcal{J}_c$ is inversely proportional to the density of states at the Fermi level $\rho\text{$_c(x$)}$. Assuming that the latter  increases monotonically with $x$ this behaviour is indicated by the concavity of the black line in Figure\,5\,(b) in the main text.

\subsection{Induced order for singlet-singlet system}
\label{app:induced}

We consider here a system with a singlet ground state $|gs\rangle$ and a singlet first excited state $|es\rangle$ with energy splitting $\Delta$, and an ordered magnetic moment along the $c$-axis as in the LMAFM phase of URu$_{2-x}$Fe$_x$Si$_2$. Singlets do not carry a magnetic moment by themselves, i.e., $\langle gs|J_z|gs\rangle=0$, so that magnetism cannot originate from the ordering of local magnetic moments that also exist above the critical temperature $T_c$, as in an ordinary magnet. However, an induced type of order may occur if there exists a non-vanishing matrix element $\alpha = \langle gs | J_z | es \rangle$ that couples the ground state and excited state singlets. The magnetic properties of the system can then be characterized by the control parameter:
\begin{equation}
    \xi = 2 \frac{\alpha^2 I_e}{\Delta}
\end{equation}
For the  derivation the reader is referred to\,\cite{Thalmeier2002, thalmeier2021}. Here we only report the main results in brief as a service to the reader. The critical temperature $T_c$ and the saturation moment $\langle  J_z  \rangle _{T=0}$ can be written as:
\begin{align}
&T_c = \frac{\Delta}{2 \tanh ^{-1} \frac{1}{\xi}} \\
&\langle  J_z  \rangle _{T=0} = \alpha \frac{1}{\xi} (\xi^2-1)^{\frac{1}{2}}
\end{align}
For $\xi>1$, a spontaneous moment can form and, for $\xi\gg1$, this moment is as large as $\alpha$, i.e. $\langle J_z  \rangle _{T=0}=\alpha$. When the two singlets in question are $\Gamma_1^{(1)}(\theta) =  \frac{1}{\sqrt{2}} \sin\theta ( | +4 \rangle + | -4 \rangle )+\frac{1}{\sqrt{2}} \cos\theta|0\rangle$ and $\Gamma_2 =\frac{1}{\sqrt{2}}  ( | +4 \rangle - | -4 \rangle )$ then we find $\alpha=4\sin\theta$, implying the material in question can have up to $\langle J_z  \rangle _{T=0}=4\sin\theta$, i.e, magnetic moments as large as $4g_J\sin\theta$ in $\mu_B$, thus accommodating even the large ordered moments of UPd$_2$Si$_2$\,\cite{Ptasiewicz1981,Shemirani1993} and UNi$_2$Si$_2$\,\cite{Lin1991}.

\section{Acknowledgements}
This research was carried out at the National Synchrotron Radiation Reseach Centre (NSRRC) in Taiwan and at PETRA\,III/DESY, a member of the Helmholtz Association HGF. Research at UC San Diego was supported by the US Department of Energy, Office of Science, Basic Energy Sciences, under Grant No. DEFG02-04-ER45105 (single crystal growth) and US National Science Foundation under Grant No. DMR-1810310 (materials characterization). A.A. and A.S. gratefully acknowledge the financial support of the Deutsche Forschungsgemeinschaft under Project No. 387555779. All authors thank H. Hess from the Institute of Nuclear Physics at the University of Cologne for providing the $\gamma$-spectra of the uranium samples and the Max Planck-POSTECH- Hsinchu Center for Complex Phase Materials for support.

%

\end{document}